\documentclass[english,aps,twocolumn]{revtex4}
\usepackage[T1]{fontenc}
\usepackage[latin9]{inputenc}
\usepackage{color}
\usepackage{graphicx}

\makeatletter

\providecommand{\tabularnewline}{\\}

\@ifundefined{textcolor}{}
{%
 \definecolor{BLACK}{gray}{0}
 \definecolor{WHITE}{gray}{1}
 \definecolor{RED}{rgb}{1,0,0}
 \definecolor{GREEN}{rgb}{0,1,0}
 \definecolor{BLUE}{rgb}{0,0,1}
 \definecolor{CYAN}{cmyk}{1,0,0,0}
 \definecolor{MAGENTA}{cmyk}{0,1,0,0}
 \definecolor{YELLOW}{cmyk}{0,0,1,0}
}

\makeatother

\usepackage{babel}
\begin{document}
\title{Slice collective dynamics, projected emittance deterioration and Free
Electron Laser performances detrimental effects}
\author{$^{1}$G. Dattoli, $^{2}$S. Di Mitri, $^{1}$F. Nguyen and $^{1}$A.
Petralia}
\affiliation{$^{1}$ENEA Fusion and Nuclear Safety Department, R.C. Frascati, 00044
Frascati, Rome, Italy}
\address{$^{2}$Elettra Sincrotrone Trieste, 34149 Basovizza, Trieste, Italy}
\begin{abstract}
The dynamical effects inducing geometrical and phase space misalignment
of bunch slice in X-ray operating Free Electron Lasers can be traced
back to a plethora of phenomena, both in the linac accelerating section
or inside the beam transport optic magnet. They are responsible for
a spoiling of the beam projected qualities and induce, if not properly
corrected, an increase of the saturation length and a decreasing of
the output power. We discuss the inclusion of these effects in models
employing scaling formulae.
\end{abstract}
\maketitle

\section{Introduction}

In forty years of research it has emerged that the physics of Free
Electron Laser (FEL) and the relevant design can be afforded by the
use of numerical codes capable of including as much as physics is
possible \citep{giannessi2003,Biedron2000}. On the other side a particular
useful role has been played by a set of semi-analytical relations,
called, perhaps improperly, scaling formulae \citep{MXie,dattoli-letardi1984,dattoli-caloi1989,BookletFEL,SaldinModel}.
This last tool has been developed, \textquotedblleft optimized\textquotedblright{}
along the course of the years and benchmarked through a wise combination
of theoretical concepts, numerical methods and comparison with the
experiment. 

The use of scaling formulae cannot replace that of dedicated codes,
but it can be helpful to fix the design working points and specify
how the various pieces of the game should be embedded. The strength
of this procedure stems from the fact that the method gathers together
the various parameters entering the definition of the FEL dynamics
by picking out those representative dimensionless global quantities
capable of quantifying, in a simple way, effects like the gain reduction,
increase of the saturation length and limitations of the output intensity.
According to refs. \citep{dattoli-letardi1984,dattoli-caloi1989,BookletFEL,Dattoli1993FELLectures},
the inhomogeneous broadening parameters are ideally suited to accomplish
such a task. They measure the deviation from the ideal beam conditions
(zero emittance, zero energy spread). The term ``inhomogeneous broadening''
traces back to the conventional laser physics. It takes into account
the broadening of the gain curve induced by the non-ideal beam qualities
(mainly energy spread, emittance, angular divergence) and its relevance
is measured with respect to the width of the \textquotedblleft natural\textquotedblright{}
line, which in the high-gain regime is associated with the Pierce
parameter. The pivotal elements of the discussion are the beam qualities,
yielding the laser non-homogeneous broadening, and the Pierce parameter,
which plays, within the FEL physics, a manifold role. The methodology,
developed within the context of the FEL scaling formulae, provides
a quantitative criterion to specify e. g. saturation length and output
power.

Further contributions, causing a dilution of the bunching mechanism
and hence of the FEL performances, may be determined by the interplay
between slippage and bunch length \citep{Dattoli1993FELLectures}.
In the case of FEL oscillators the detrimental effects associated
with the lack of longitudinal overlapping and induced mode locking
are well documented either theoretically and experimentally \citep{ColsonLaserHand}.
Regarding the high-gain SASE FEL regime, operating in the X-ray region,
the effect is even more rich and has provided a wealth of new phenomenology,
involving the contributions to the lasing process due to slice and
projected emittances.

The electron beam transport has acquired new perspectives and has
opened new problems to be solved. Each slice is characterized by its
own phase space distribution and a good alignment, along the electron
bunch, is the prerequisite for good performances for short wavelengths
(X-ray) FELs \citep{Tanaka2014,Dattoli2004,Dattoli2012,DiMitri2014,guert2015}.
This effect is even more delicate than it may appear, since it involves
geometrical and phase space alignments . For example, a transverse
shift or tilt, even though leaving unaltered the slice emittance,
may be a spoiling source for the projected counterpart and for the
associated transport optics \citep{Dattoli2004,DiMitri2014}. 

In this paper we use the concepts and the formalism developed in \citep{dattoli-letardi1984,dattoli-caloi1989,BookletFEL,Dattoli2012}
and make the attempt of framing, within the context of models developed
in \citep{Dattoli2004,DiMitri2014}, these bunching smearing effects.
The forthcoming section is devoted to a description of the procedure
we intend to use and to the relevant implementation to include the
slice (transverse) misalignment and tilt contributions to the increase
of e.g. saturation length. Section three is finally devoted to comparison
with numerical results and final comments. 

\section{Scaling formulae and slice tilting}

The inhomogeneous parameters associated with the electron beam transverse
dimensions and divergences are specified by the identities

\begin{equation}
\begin{array}{l}
\tilde{\mu}_{\eta'}=\frac{\varepsilon_{\eta}}{\beta_{\eta}}\frac{\gamma^{2}}{\rho}\frac{1}{\left(1+\frac{K^{2}}{2}\right)}\\
\tilde{\mu}_{\eta}=\varepsilon_{\eta}\frac{\gamma^{2}}{\rho}\frac{1}{\left(1+\frac{K^{2}}{2}\right)}\frac{\left(\gamma_{T}^{*}\right)^{2}}{\gamma_{\eta}}
\end{array}\label{eq:1}
\end{equation}

\noindent where $\eta$ indicates the transverse dimensions $x,y$,
$\gamma$ is the electron relativistic factor, $\lambda_{u}$ the
undulator period and $B_{0}$ the peak of the on-axis magnetic field.
$\beta_{\eta},\gamma_{\eta}$ are the matching Twiss parameters while
with $\beta_{\eta}^{*},\gamma_{\eta}^{*}$ we denote the Twiss parameters
for matching to the undulator natural focusing (NFTP). $a_{w}=K/\sqrt{2}$
for a for planar undulators and $a_{u}=K$ for helical undulators
where $K=\frac{\lambda_{u}\left[cm\right]B_{0}[KG]}{10.71}$ is the
undulator strength parameter. 

\noindent The Pierce parameter $\rho$, expressed in practical units,
is

\begin{equation}
\begin{array}{l}
\rho\cong\frac{8.36\cdot10^{-3}}{\gamma}\left[J\left(\frac{A}{m^{2}}\right)\left(\lambda_{u}(cm)\,Kf_{b}(K)\right)^{2}\right]^{\frac{1}{3}}\\
J=\frac{I}{2\pi\sigma_{x}\sigma_{y}}\\
\sigma_{\eta}=\sqrt{\beta_{\eta}\varepsilon_{\eta}}
\end{array}\label{eq:2}
\end{equation}

\noindent being $J$ the current density, with $I$ the bunch current
and $\sigma_{\eta}$ the bunch transverse rms size ($\eta=x,y$),
and $f_{b}$ is the Bessel factor ($f_{b}=1$ for helical undulator,
$f_{b}=J_{0}\left(\xi\right)-J_{1}\left(\xi\right)$ for linear undulator,
with $J_{0}$ and $J_{1}$ the Bessel functions of order 0 and $1$,
and $\xi=\frac{1}{4}K^{2}{\left(1+\frac{K^{2}}{2}\right)}^{-1}$).
From here on the case of a linear undulator (hence $a_{w}=K/\sqrt{2}$)
will be considered but the treatment remain valid also for a helical
undulator.

\noindent We consider in the following that the average transverse
sizes of the beam in the two radial and vertical directions are the
same in the interaction region, ${\sigma}_{x}={\sigma}_{y}$ . This
condition is realistic in an undulator magnetic lattice by imposing,
as matching condition, that the difference between the x and y size
of the beam has to be minimized \citep{quattromini2012}. In addition
we assume a beam with identical transverse emittances in both radial
and vertical directions so that ${\varepsilon}_{x}={\varepsilon}_{y}$.

\noindent For this reason from eq.(\ref{eq:2}) we can define a value
of emittance $\varepsilon$, beam size $\sigma$ and for the Twiss
parameter ${\beta}_{\eta},{\alpha}_{\eta},{\gamma}_{\eta}$ in such
a way that

\begin{equation}
\begin{array}{l}
{\varepsilon}_{x}={\varepsilon}_{y}=\varepsilon\\
{\sigma}_{x}={\sigma}_{y}=\sigma\\
{\beta}_{x}={\beta}_{y}={\beta}_{T}\\
{\gamma}_{x}={\gamma}_{y}={\gamma}_{T}\\
{\alpha}_{x}={\alpha}_{y}={\alpha}_{T}
\end{array}\label{eq:3}
\end{equation}

\noindent As we will see in the following, a crucial role is played
by $\beta_{T},\gamma_{T}$ and by the NFTP linked to the undulator
period and strength by the relations
\begin{equation}
\gamma_{T}^{*}=\frac{1}{\beta_{T}^{*}},\ \beta_{T}^{*}=\frac{\gamma\,\lambda_{u}}{\pi\,K}.\label{eq:4}
\end{equation}

\noindent Before proceeding further let us note that, from eq.(\ref{eq:2})
and with the assumpion in eq.(\ref{eq:3}), the Pierce parameter exhibits
the following dependence on the beam current density 
\begin{equation}
\rho\propto J^{\frac{1}{3}}=\left(\frac{I}{2\,\pi\,\sigma^{2}}\right)^{\frac{1}{3}}\label{eq:5}
\end{equation}
Therefore we define 
\begin{equation}
\rho=\rho^{*}\left(\frac{\beta_{T}^{*}}{\beta_{T}}\right)^{\frac{1}{3}}\label{eq:6}
\end{equation}
with $\rho^{*}$ being the Pierce parameter calculated with a current
density corresponding to the NFTP.

\noindent It is convenient to write eq.(\ref{eq:1}) in the more useful
form 
\begin{equation}
{\color{red}\begin{array}{l}
{\normalcolor {\tilde{\mu}_{x'}=\tilde{\mu}_{x'}^{*}\left(\frac{\beta_{T}^{*}}{\beta_{T}}\right)^{\frac{2}{3}}}}\\
{\normalcolor {\tilde{\mu}_{x}=\tilde{\mu}_{x'}^{*}\left(\frac{\beta_{T}^{*}}{\beta_{T}}\right)^{-\frac{2}{3}}\frac{1}{\beta_{T}^{*}\gamma_{T}}=\frac{\tilde{\mu}_{x'}^{*}}{1+\alpha_{T}}\left(\frac{\beta_{T}^{*}}{\beta_{T}}\right)^{-\frac{2}{3}}}}\\
{\normalcolor {\tilde{\mu}_{x'}^{*}=\frac{\varepsilon}{\beta_{T}^{*}}\frac{\gamma^{2}}{\rho^{*}}\frac{1}{\left(1+\frac{K^{2}}{2}\right)}=\frac{\varepsilon}{\lambda_{u}}\frac{\gamma^{2}}{\rho^{*}}\frac{\pi\,K}{\left(1+\frac{K^{2}}{2}\right)}=\frac{\varepsilon}{2\lambda}\frac{\pi K}{\rho^{*}}}}
\end{array}}\label{eq:7}
\end{equation}
where $\tilde{\mu}_{x'}^{*}$ it is the NFTP inhomogeneous broadening
coefficient and by remembering that the FEL resonance wavelength is
expressed by

\noindent 
\begin{equation}
\lambda=\frac{\lambda_{u}}{2\gamma^{2}}\left(1+\frac{K^{2}}{2}\right).\label{eq:8}
\end{equation}

\noindent It is worth noting from eq.(\ref{eq:1}) that $\tilde{\mu}_{x}^{*}=\tilde{\mu}_{x'}^{*}$
and that it can be expressed also in the form

\begin{equation}
{\color{red}\begin{array}{l}
{\normalcolor {\tilde{\mu}_{x}^{*}=\frac{\pi\,\varepsilon_{n}}{\lambda_{u}\rho^{*}}\phi(K),}}\\
{\normalcolor {\phi(K)=\frac{K}{1+\frac{K^{2}}{2}},}}
\end{array}}\label{eq:9}
\end{equation}

\noindent The last equations, where ${\varepsilon}_{n}=\gamma\,\varepsilon$
is the ``normalized emittance'', reveal the physical nature of these
parameters, whose meaning traces back to the inhomogeneous broadening
induced by the angular content of the beam and by its transverse dimension.
Namely the ratio of the laser line broadening to the natural width.

\noindent The request that they have to be less than unity, to avoid
problems like the increase of the saturation length, yields a condition
on emittance, namely 
\begin{equation}
{\color{red}{\normalcolor \varepsilon_{n}<\rho^{*}\frac{\lambda_{u}}{\pi\,\phi(K)}}}\label{eq:10}
\end{equation}
Which, if used along with $\varepsilon_{n}\cong\frac{\gamma\lambda}{4\,\pi}$,
yields the further constraint 
\begin{equation}
{\color{red}{\normalcolor K<8\,\gamma\rho^{*}}}\label{eq:11}
\end{equation}
The coefficients $\tilde{\mu}_{x',x}$in eq.(\ref{eq:7}) contain
the corrections to the inhomogeneous broadening due to a matching
different than that of the natural undulator focusing. It should be
noted that the effect induced by $\widetilde{\mu}_{x}$ or $\widetilde{\mu}_{x'}$
goes in opposite directions with varying $\beta_{T}$. In the following
we consider an undulator focusing in both transverse planes. This
assumption simplifies the formalism, since there is no difference
between radial and vertical $\widetilde{\mu}$ parameters.

\noindent A further quantity contributing to the bunnching smearing
due to non ideal beam qualities is the relative energy spread $\sigma_{\varepsilon}$
whose role can be quantified through the parameter 
\begin{equation}
\tilde{\mu}_{\varepsilon}=2\,\frac{\sigma_{\varepsilon}}{\rho}\label{eq:12}
\end{equation}
The FEL gain length is expressed in terms of the undulator period
and of the Pierce parameter $\rho$ as 
\begin{equation}
L_{g}=\frac{\lambda_{u}}{4\,\pi\,\sqrt{3}\rho}\label{eq:13}
\end{equation}
One of the macroscopic consequences of the inhomogeneous broadening
is that of increasing the gain length and thus the saturation length.
This effect is obtained by replacing $\rho$ with 
\begin{equation}
\begin{array}{l}
\rho_{3D}=\chi^{-1}\rho\\
L_{g3D}=\chi\,L_{g}\\
{\normalcolor {\normalcolor {\color{red}{\normalcolor \chi=F_{2}/F_{1}}}}}\\
F_{1}=\frac{1}{\sqrt{\left(1+\tilde{\mu}_{x}^{2}\right)\left(1+\tilde{\mu}_{x'}^{2}\right)\left(1+\tilde{\mu}_{y}^{2}\right)\left(1+\tilde{\mu}_{y'}^{2}\right)}}\\
F_{2}=1+0.185\,\frac{\sqrt{3}}{2}\,F_{1}\,\tilde{\mu}_{\varepsilon}^{2}\\
{d=0.185\,\frac{\sqrt{3}}{2}}
\end{array}\label{eq:14}
\end{equation}
We refers to this last parametrization as the DOP model (from the
name of the authors G.Dattoli, P.L.Ottaviani, S.Pagnutti \citep{BookletFEL})
and predictions are in agreement with those from the Xie/Saldin models
\citep{MXie,SaldinModel}.

\begin{figure}
\includegraphics[width=0.8\columnwidth]{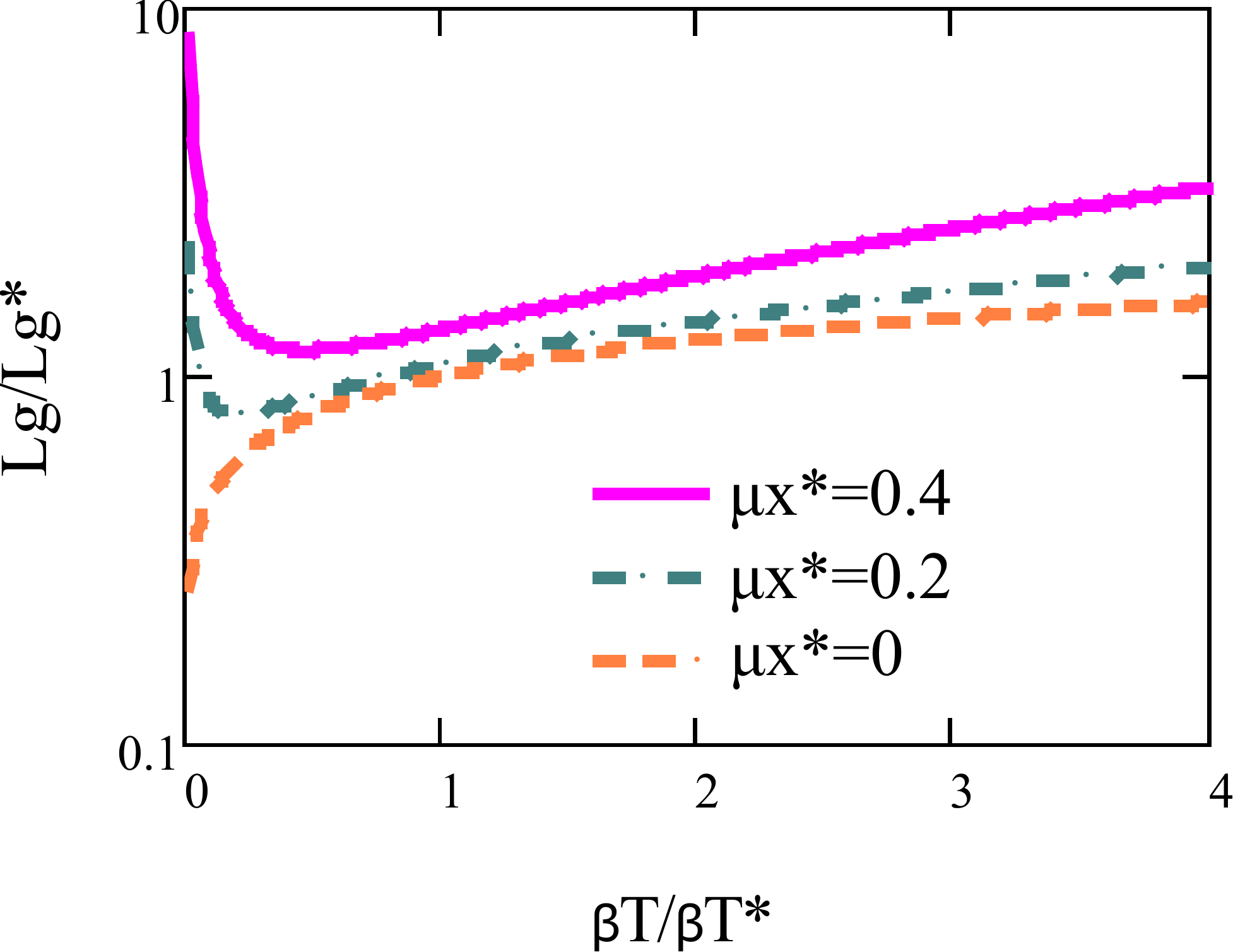}\caption{Gain length as a function of the Twiss $\beta$ parameter, from eq.(\ref{eq:15}).
Both quantities are given in terms of their counterparts evaluated
for the case of natural focusing for the undulator (indicated with
the star symbol). For this reason the ratio $L/L_{g}^{*}$ is plotted
(in logaritmic scale) as a function of the ratio $\beta_{T}/\beta_{T}^{*}$
in the case with null $\alpha_{T}$ Twiss matching parameter, for
different value of $\widetilde{\mu}_{x}^{*}$. The case $\widetilde{\mu}_{x}^{*}=0$
corresponds to a null emittance (see eq.(\ref{eq:9})) hence to no
inhomogeneous broadening effects. }

\label{gainlengthbeta}
\end{figure}

\noindent Assuming a round beam and equal focusing properties of the
undulator the $\tilde{\mu}$ coefficients are the same in both planes
($\tilde{\mu}_{x}=\tilde{\mu}_{y}$), in the matching condition with
$\alpha_{T}=0$, and neglecting the effect of the energy spread we
can write the gain length as 
\begin{equation}
\begin{array}{l}
{L_{g}=L_{g}^{*}X^{\frac{1}{3}}\left(1+\tilde{\mu}_{x}^{*2}X^{-\frac{4}{3}}\right)\left(1+\tilde{\mu}_{x}^{*2}X^{\frac{4}{3}}\right),}\\
{X=\frac{\beta_{T}}{\beta_{T}^{*}}}
\end{array}\label{eq:15}
\end{equation}

\noindent where $L_{g}^{*}=\frac{\lambda_{u}}{4\pi\sqrt{3}\,\rho*}$
is the gain length evaluated in the condition of natural focusing
of the undulator (see eqs.(\ref{eq:4}),(\ref{eq:6})). The expression
in eq.(\ref{eq:15}) is plotted in Fig.\ref{gainlengthbeta}. In absence
af any inhomogenous broadening effectsThe case $\widetilde{\mu}_{x}^{*}=0$,
corresponding to a null emittance (see eq.(\ref{eq:9})), gives the
case 

\noindent 
\noindent Albeit this procedure is not entirely correct, for the reasons
discussed in the concluding section, the conclusion we draw from eq.(\ref{eq:15})
regarding the \textquotedbl optimum\textquotedbl{} beta are sufficiently
accurate to be considered reliable.

\noindent The evolution of the FEL power can be expressed in terms
of a function of logistic nature 
\begin{equation}
\begin{array}{l}
{P\left(z\right)=P_{0}\frac{A\left(z\right)}{1+\frac{P_{0}}{P_{F}}\left[A\left(z\right)-1\right]}}\\
{A\left(z\right)=\frac{1}{9}\left[3+2\cosh\left(\frac{z}{L_{g}}\right)+4\cos\left(\frac{\sqrt{3}}{2}\frac{z}{L_{g}}\right)\cosh\left(\frac{z}{2L_{g}}\right)\right]}
\end{array}\label{eq:16}
\end{equation}
where $P_{F}=\sqrt{2}\rho P_{E}$ is the FEL saturation power and
$P_{E}$ is the electron beam energy, in which $L_{g}$contains the
induced non homogeneous effects through the redefinition of the Pierce
parameter. The effect of the beam qualities on the output power will
not be discussed in this paper therefore the $\rho$ parameter, defining
the saturated laser power $P_{F}$, does not include any correction.

\section{Emittance and collective effects}

\noindent A further element of complexity is provided by the role
played by slice and projected emittances. The distinction arises when
the slippage length (namely the mismatch between laser and electrons
due to the different velocities) is significantly shorter than the
bunch length.

\noindent Assuming that slippage and coherence length $l_{c}$ coincide,
we have 
\begin{equation}
l_{s}\cong l_{c}\cong\frac{\lambda}{4\pi\sqrt{3}\rho}\label{eq:17}
\end{equation}
being $\lambda$ the resonanche wavelength of the FEL radiation (\ref{eq:8}).
In literature different definitions of $l_{c}$ can be found, being
limited to numerical consants but they do not produce any significan
deviations regarding the physical consequences. The number of slices
is approximately fixed by the ratio of the bunch length to the slippage/coherence
length, namely 
\begin{equation}
m\cong\frac{\sigma_{z}}{l_{c}}\label{eq:18}
\end{equation}
The slices grows almost independently, each of them is characterized
by their own phase space, emittance and energy spread. Here we use
the coasting beam approximation and assume that each slice has identical
phase space distribution, emittance and energy spread.

\noindent The projected beam qualities are all referred to the whole
bunch. Projected and slice phase spaces are characterized by different
dynamics. The collective effects, as CSR and GTW, may create misalignments
increasing the projected emittance and leaving unaltered the slice
counterparts \citep{DiMitri2014}. In other words the projected emittance
is expected to be larger than the slice one.

\noindent The point we like to raise is how to account for the detrimental
effects induced on the FEL dynamics by the increase of the projected
emittance using the criteria developed in the previous picture.

\noindent Previous papers have addressed this problem. In particular,
in \citep{Dattoli2012} it has been afforded by defining the projected
emittance by including the different phase space distributions of
the individual slices and calculating the emittance growth including
the statistical effects deriving from the Twiss coefficients characterizing
the individual slices.

\noindent In \citep{DiMitri2014} a different analysis has been performed
and it has been shown that a $\tilde{\mu}$-like parameter can be
defined and the increase of gain length due to the induced growth
of the projected emittance can be naturally included by embedding
it within a procedure much similar to that discussed so far.

\noindent We have already noted that emittance contributes to the
bunching smearing through the (incoherent) contributions due the angular
and spatial contents of the bunch distribution, the coherent (collective)
contribution is associated with the rms divergence $\left\langle \vartheta^{2}\right\rangle $
of the slices centroids inside the undulator. The dilution of the
bunching due to this contribution may become even larger than that
corresponding to the incoherent parts.

\noindent The reference parameter adopted in \citep{DiMitri2014}
is 
\begin{equation}
\begin{array}{l}
{\tilde{\mu}_{coll}=\frac{\left\langle \vartheta^{2}\right\rangle }{\vartheta_{0}^{2}},}\\
{\vartheta_{0}^{2}=\frac{\lambda}{L_{g}}}
\end{array}\label{eq:19}
\end{equation}
where the subscript ``coll'' stands for ``collective''. The critical
angle $\vartheta_{0}$, in terms of the FEL characteristic quantities,
can be written as 
\begin{equation}
\vartheta_{0}^{2}=\frac{\lambda}{L_{g}}=\frac{2\pi\sqrt{3}\rho}{\gamma^{2}}\left(1+\frac{K^{2}}{2}\right)\label{eq:20}
\end{equation}
Thus getting 
\begin{equation}
\tilde{\mu}_{coll}=\frac{\gamma^{2}\left\langle \vartheta^{2}\right\rangle }{\left(1+\frac{K^{2}}{2}\right)2\pi\sqrt{3}\rho}\label{eq:21}
\end{equation}
The analogy with $\tilde{\mu}_{x'}$ is evident. If we interpret $\frac{\varepsilon}{\beta_{T}}=\left\langle x'^{2}\right\rangle $
as beam divergence we can write $\tilde{\mu}_{x'}$ as 
\begin{equation}
\tilde{\mu}_{x'}=\frac{\varepsilon}{\beta_{T}}\frac{\gamma^{2}}{\rho}\frac{1}{\left(1+\frac{K^{2}}{2}\right)}=\frac{\gamma^{2}\left\langle x'^{2}\right\rangle }{\rho\left(1+\frac{K^{2}}{2}\right)}=2\pi\sqrt{3}\frac{\left\langle x'^{2}\right\rangle }{\vartheta_{0}^{2}}\label{eq:22}
\end{equation}
The physical nature of $\tilde{\mu}_{coll}$ can therefore be considered
not extraneous to the previously outlined formalism.

\noindent In ref. \citep{DiMitri2014} it has been proposed and checked,
by comparison with computer simulation, that such an inclusion occurs
through the following re-definition of the $\chi$ function

\begin{equation}
L_{g,coll}=\frac{\chi}{1-\varsigma\,\tilde{\mu}_{coll}\,\chi}\,L_{g}\label{eq:23-Lgcoll-tanaka}
\end{equation}
which will be commented in the following (if $\zeta=\pi$ the previous
equation reduces to the Tanaka formula derived in ref.\citep{Tanaka2014}). 

\noindent The previous correction holds for the saturation length
associated with detrimental effects due to slice tilting inside the
electron bunch and eventually to the increase of the emittance to
the projected value 

\noindent 
\begin{equation}
\varepsilon_{coll}=\varepsilon\sqrt{\left|\begin{array}{cc}
{\beta_{T}} & {\alpha_{T}}\\
{-\alpha_{T}} & {\gamma_{T}+\frac{\left\langle x_{coll}'^{2}\right\rangle }{\varepsilon}}
\end{array}\right|}\label{eq:24}
\end{equation}
with the $\gamma_{T}$ Twiss coefficient modified by the inclusion
of the slice centroids associated divergence. Assuming $\alpha_{T}=0$
we express the emittance growth due to collective effcts as 
\begin{equation}
\varepsilon_{coll}=\varepsilon\sqrt{1+\frac{\beta_{T}\left\langle x_{coll}'^{2}\right\rangle }{\varepsilon}}\label{eq:25}
\end{equation}

and hence the normalized collective (projected) emittance $\epsilon_{n,coll}=\gamma\epsilon_{coll}$. 

\begin{table}[h]
\begin{tabular}{|l|c|c|}
\hline 
Parameter & Value & Unit\tabularnewline
\hline 
\hline 
Energy $E$ & 1.8 & GeV\tabularnewline
\hline 
Peak current $I_{p}$ & 3 & kA\tabularnewline
\hline 
Norm. slice emittance $\epsilon_{n}$ & 0.5 & mm mrad\tabularnewline
\hline 
Norm. collective projected emittance $\epsilon_{n,coll}$  & 2.3 & mm mrad\tabularnewline
\hline 
Undulator parameter (planar und.) $K$ & $\sqrt{2}$ & \tabularnewline
\hline 
Undulator period $\lambda_{u}$ & 20 & mm\tabularnewline
\hline 
Resonance wavelength $\lambda$ & 1.6 & nm\tabularnewline
\hline 
\end{tabular}

\caption{\label{tab:beampar}Set of parameters used to compare the gain length
modified by collective effects with the $L_{g,3D}$, as a function
of the Twiss $\beta$ parameter. The slice emittance is supposed to
be the same for all slices. Projected emittance coincides with the
projected emittance if no misaliglments of the slices occur in the
phase space.}
\end{table}
\begin{figure}[h]
\includegraphics[width=0.9\columnwidth]{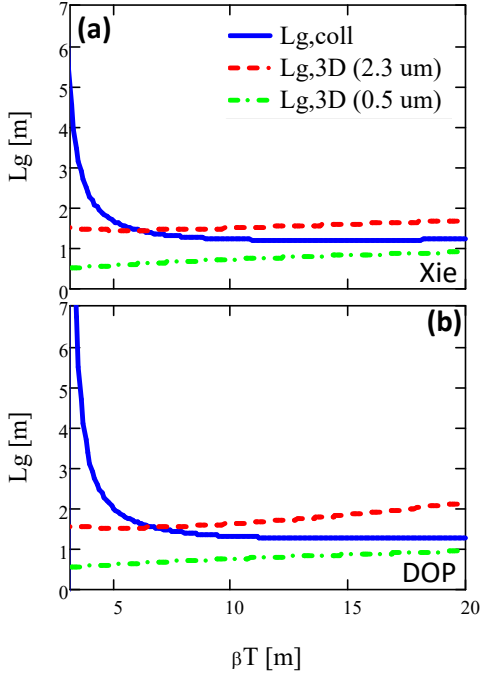}\caption{\label{fig2}\textcolor{red}{{} }Gain length, calculated with paralmeters
listed in Tab.\ref{tab:beampar}, as a function of the average Twiss
$\beta$ parameter in the undulator according to eq.(\ref{eq:23-Lgcoll-tanaka}).
In (a) $L_{g,3D}$ has been evaluated according to the Xie formalism
according to results in \citep{DiMitri2014}, in (b) the DOP model,
used in the present paper, has been considered to evaluate the ihmonogeneous
broadening effects on the gain length. Dash green and dot red lines
refer to situations when no misaligment is present, in these cases
the projected emittance coincides with the slice emittance and $L_{g,coll}$
reduces to $L_{g,3D}$, the two lines are obtained respectively for
$\epsilon_{n}=0.5\,mm\,mrad$ and $\epsilon_{n}=2.3\,mm\,mrad$. The
blue continuous line refers to $L_{g,coll}$ in presence of a misaligment
where the norm. slice emittance $\epsilon_{n}=0.5\,mm\,mrad$ and
the norm. projected value is $\epsilon_{n,coll}=2.3\,mm\,mrad$. In
all cases the effects of the energy spread have not been considered
($\widetilde{\mu}_{\epsilon}=0)$. It is evident that for small $\beta_{T}$
values an increasing of the gein length occurs, being this conditon
of course inauspicious for FEL operation.}
\end{figure}
\noindent The previous identity becomes a measure of $\left\langle x'^{2}\right\rangle $
through the evaluation of the projected emittance. We set

\noindent 
\begin{equation}
\begin{array}{l}
\left\langle x'^{2}\right\rangle =\Delta\frac{\epsilon}{\beta_{T}}\\
\Delta=\left(\frac{\epsilon_{coll}}{\epsilon}\right)^{2}-1
\end{array}\label{eq:26}
\end{equation}
and write $\widetilde{\mu}_{coll}$ as

\noindent 
\begin{equation}
\widetilde{\mu}_{coll}=\frac{\Delta}{\beta_{T}}\frac{\lambda_{u}}{4\pi\sqrt{3}\rho}\frac{\epsilon}{\lambda}\label{eq:27}
\end{equation}
According to the previous identity, large beta values should smear
out the detrimental effect of the collective tilt. 

\noindent The use of the parameter $X$, introduced in eq.(\ref{eq:15})
allows to cast eq.(\ref{eq:27}) in the more convenient form 

\noindent 
\begin{equation}
\widetilde{\mu}_{coll}=\frac{\Delta}{\beta_{T}^{*}}\frac{\lambda_{u}}{4\pi\sqrt{3}\rho^{*}X^{2/3}}\frac{\epsilon}{\lambda}\label{eq:28}
\end{equation}
The correction to the saturation length foreseen in eq.(\ref{eq:23-Lgcoll-tanaka}),
predicts a critical value of $\beta_{T}$ for which the collective
effects are dominating over those associated with those due to the
inhomogeneous broadening. This critical value occurs when in eq.(\ref{eq:23-Lgcoll-tanaka})
$\widetilde{\mu}_{coll}=\zeta^{-1}$, namely when

\noindent 
\begin{equation}
\beta_{ct}=\left[\frac{\Delta}{\zeta\beta_{T}^{*}}L_{g}^{*}\frac{\epsilon}{\lambda}\right]^{3/2}\beta_{T}^{*}\label{eq:29}
\end{equation}
obtained by assuming negligible inhomogeneous effects due to emittance,
choosing $\frac{\epsilon}{\lambda}=\frac{1}{4\pi}$ eventually yields

\noindent 
\begin{equation}
\beta_{ct}=\left[\frac{\Delta}{4\pi\zeta\beta_{T}^{*}}L_{g}^{*}\right]^{3/2}\beta_{T}^{*}\label{eq:30}
\end{equation}
In correspondence of this value the saturation length becomes prohibitively
long, at least according to the assumption that $\widetilde{\mu}_{coll}$
induced effect are accounted by an equation of the type (\ref{eq:23-Lgcoll-tanaka}).
In Fig.\ref{fig2} we have compared the saturation length vs. $\beta_{T}$
for the the cases corresponding to a normalized slice emittance $\epsilon_{n}$,
a normalized projected emittance $\epsilon_{n,coll}$ and to the collective
effect with parameters of Tab.\ref{tab:beampar} for the electron
beam and radiation. Very similar results are obtained by using the
Xie (used also in \citep{DiMitri2014}) or the DOP formalism to evaluate
the 3D corrections to the gain length. The plot axes are limited to
a minimum considered $\beta_{T,min}=3$ because (with the assumed
parameters for the electron beam) a singulariy occurs, due to the
form of eq.(\ref{eq:23-Lgcoll-tanaka}), for lower $\beta_{T}$ values
where the equation is supposed to have no more physical sense. It
is worth noting that when $\beta_{T}\gg\beta_{ct}$ the saturation
length keeps reasonable values, in between those associated with slice
and collective emittance. 

\section{An alernative point of view}

\noindent The contribution of the collective effects, using eq.(\ref{eq:23-Lgcoll-tanaka}),
appears as a separate function, appended to the inhomogeneous $\chi$
function and without affecting the other quantities entering the definition
of the other $\widetilde{\mu}$ parameters.

\noindent A different way of looking at the increasing of the gain
length caused by the collective effects, is to leave the same expression
as in eqs.(\ref{eq:14}), but replacing $\varepsilon$ with $\varepsilon_{coll}$
wherever the emittance appears.

\noindent Regarding the Pierce parameter and the fact that it is proportional
to the cubic root of the beam current density, the correction due
to the ``collective'' emittance yields 
\begin{equation}
\begin{array}{l}
{\rho_{coll}=\rho\left(\frac{\sigma}{\sigma_{coll}}\right)^{\frac{2}{3}}=\rho\left(\frac{\epsilon}{\epsilon_{coll}}\right)^{\frac{1}{3}},}\\
{\sigma_{coll}=\sqrt{\beta_{T}\varepsilon_{coll}}}
\end{array}\label{eq:31}
\end{equation}
While as to the inhomogeneous broadening we end up with 
\begin{equation}
\begin{array}{l}
{\tilde{\mu}_{x',coll}=\frac{\varepsilon_{coll}}{\beta_{T}}\frac{\gamma^{2}}{\rho_{coll}}\frac{1}{\left(1+\frac{K^{2}}{2}\right)},}\\
{\tilde{\mu}_{x,coll}=\varepsilon_{coll}\frac{\gamma^{2}}{\rho_{coll}}\frac{1}{\left(1+\frac{K^{2}}{2}\right)}\frac{\left(\gamma_{T}^{*}\right)^{2}}{\gamma_{T}},}\\
{\tilde{\mu}_{\varepsilon,coll}=\frac{2\,\sigma_{\varepsilon}}{\rho_{coll}}}
\end{array}\label{eq:32}
\end{equation}
It is more convenient to cast them in the form 
\begin{equation}
\begin{array}{l}
{\tilde{\mu}_{x',coll}=\tilde{\mu}_{x'}\left(\frac{\varepsilon_{coll}}{\varepsilon}\right)^{\frac{5}{3}},}\\
{\tilde{\mu}_{x,coll}=\tilde{\mu}_{x}\left(\frac{\varepsilon_{coll}}{\varepsilon}\right)^{\frac{5}{3}}}\\
{\tilde{\mu}_{\varepsilon,coll}=\tilde{\mu}_{\varepsilon}\left(\frac{\varepsilon_{coll}}{\varepsilon}\right)^{\frac{1}{3}}}
\end{array}\label{eq:33}
\end{equation}
which shows that the correction associated with slice misalignment
can be characterized by the ratio $\varepsilon_{coll}/\varepsilon$.

\begin{figure}
\includegraphics[width=0.7\columnwidth]{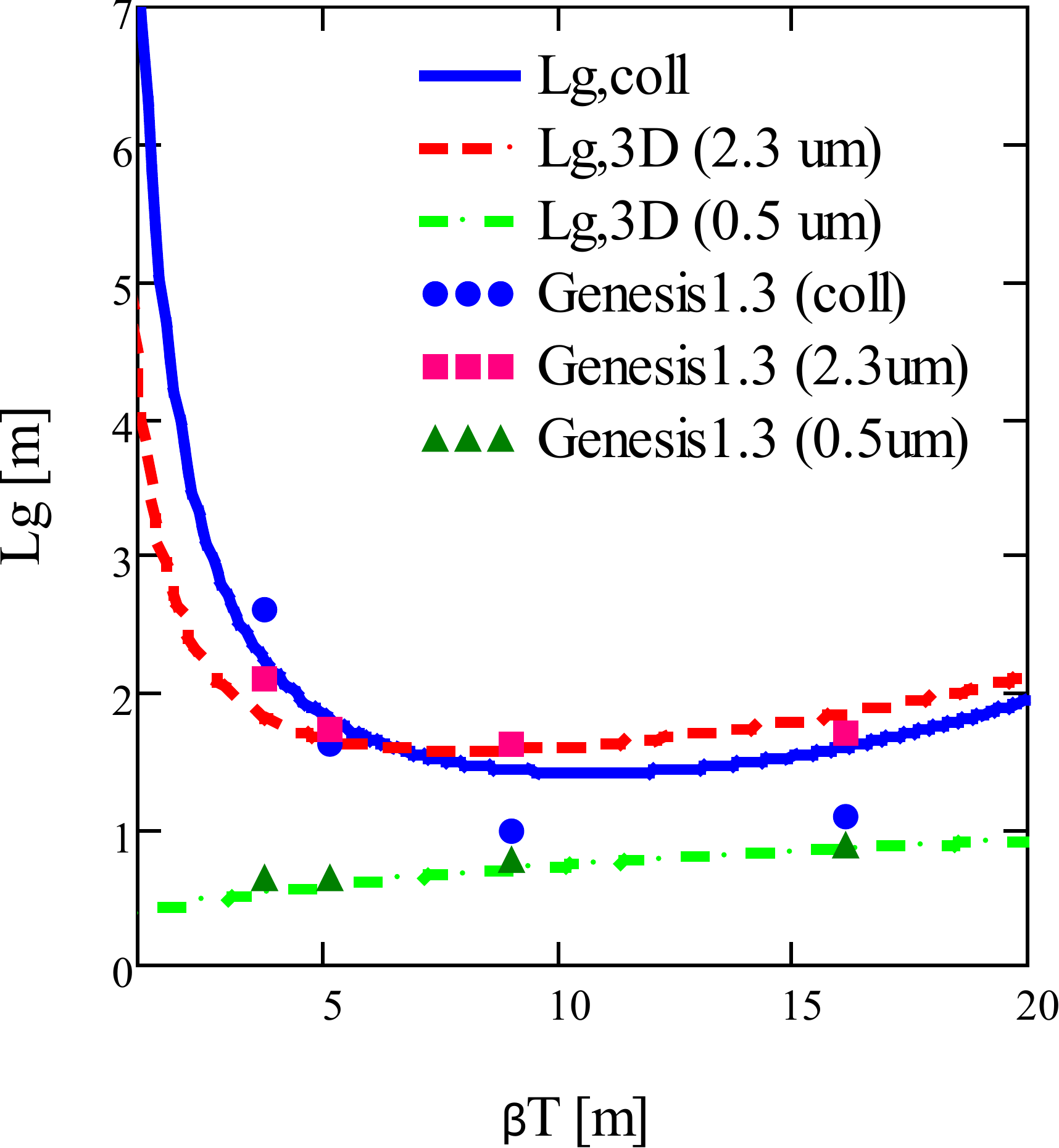}

\caption{\label{fig:Lg-vs-Genesis} Gain length with the inclusion of the collective
effects evaluated from equation (\ref{eq:34-Lcoll-alt}). Dash green
and dot red lines refer to situations when no misaligment is present
and respectively for $\epsilon_{n}=0.5\,mm\,mrad$ and $\epsilon_{n}=2.3\,mm\,mrad$.
The blue continuous line is the gain length $L_{g}^{(coll)}$with
the inclusion of collective effectst with a norm. slice emittance
$\epsilon_{n}=0.5\,mm\,mrad$ and a norm. projected value $\epsilon_{n,coll}=2.3\,mm\,mrad$.
Energy spread effects are not accounted ($\widetilde{\mu}_{\epsilon}=0)$.
The formula fits well with a set of GENESIS 1.3 numerical simulations,
described in ref.\citep{DiMitri2014}, expecially in the region around
the minimum $\beta_{T}$ in the optimum range of operation for a FEL
with parameters of Tab.\ref{tab:beampar}.}
\end{figure}

With these redefinition of the $\widetilde{\mu}$ parameters, the
$\chi$ function of eq.(\ref{eq:14}) can be rewritten in order to
achieve a corrected albeit approximate expression wich includes the
collective effects. We can write the function of the $\rho$ depleting
formula, and hence the gain length as 
\begin{equation}
\frac{L_{g}^{(coll)}}{L_{g}}=\chi_{coll}\cong1+\left(\frac{\varepsilon_{coll}}{\varepsilon}\right)^{\frac{2}{3}}\left[\left(\tilde{\mu}_{x}^{2}+\tilde{\mu}_{x'}^{2}\right)\left(\frac{\varepsilon_{coll}}{\varepsilon}\right)^{\frac{8}{3}}+d\tilde{\mu}_{\varepsilon}^{2}\right]\label{eq:34-Lcoll-alt}
\end{equation}
obtained by neglecting the cross products, where it is also evident
the increase of the gain length (hence of the saturation length) at
low $\beta_{T}$ values. (\ref{eq:34-Lcoll-alt}) has been ploted
in Fig.\ref{fig:Lg-vs-Genesis} and compared with a set of numerical
simulations, already shown in ref.\citep{DiMitri2014} and made with
the code GENESIS 1.3 \citep{genesis1.3}, with good agreement expecially
for $\beta_{T}$ values in the reasonable range of operation for an
actual FEL with parameters in Tab.\ref{tab:beampar}. 

\begin{figure}
\includegraphics[width=0.8\columnwidth]{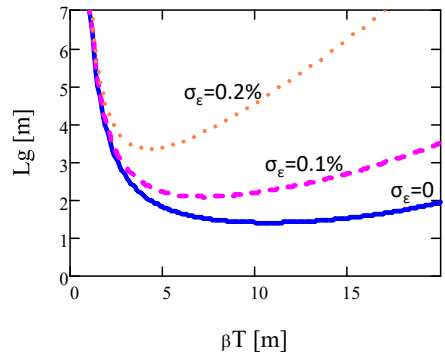}

\caption{\textcolor{red}{\label{fig:Lgcoll-complete}}Gain length $L_{g}^{(coll)}$
calculated with paralmeters listed in Tab.\ref{tab:beampar}, as a
function of the average Twiss $\beta_{T}$ parameter in the undulator,
with the inclusion of the effets due to energy spread $\sigma_{\epsilon}=2\rho\widetilde{\mu}_{\epsilon}$.
Different lines have been calculated, with a norm. slice emittance
$\epsilon_{n}=0.5\,mm\,mrad$ and a norm. collective projected emittance
$\epsilon_{n,coll}=2.3\,mm\,mrad$, for different values of the energy
spread. Blue continuous, dash purple and dot orange lines are respectively
for $\sigma_{\epsilon}=0$, $\sigma_{\epsilon}=0.1\%$ and $\sigma_{\epsilon}=0.2\%$.
It is evident, as expected, a deterioration of the beam, and hence
an increasing of the gain length, with higher energy spread values. }
\end{figure}

\noindent In Fig.\ref{fig:Lgcoll-complete} we have plotted the gain
length $L_{g}^{(coll)}$ (eq.(\ref{eq:34-Lcoll-alt})) vs. the Twiss
parameter beta with also including the effect of the energy spread.
The results show that the optimum $\beta_{T}$ value depends on the
beam qualities. It is worth stressing that in the presence of a larger
energy spread the optimization procedure may become critical.


\section{Final Comments}

In this paper we have addressed the problem of including the increase
of the projected emittance growth due to slice misalignment, in the
evaluation of the SASE FEL gain. Thus determining the relevant detrimental
effects on the saturation length. The key element of the analysis
is the introduction of an inhomogeneous broadening like parameter
\citep{DiMitri2014} which measures the increase of emittance due
to collective effects. In order to preserve a line of continuity with
the treatment in refs.\citep{MXie,SaldinModel} it is assumed that
this effect has not separated function and affects the saturation
length through a suitable redefinition of the Pierce parameter. An
effective idea of the interplay between the various detrimental contributions
to the the increase of the saturation length is offered by the approximate
expression of the $\chi$ function given in eq.(\ref{eq:34-Lcoll-alt})
which yields an idea of how the various terms due to collective, slice
emittances and energy spread contribute to the increase of the gain
length. All of the two models indicates that a strong deterioration
of the gain length occurs al low values of the matching $\beta_{T}$
Twiss parameter where therefore the operation of the FEL is not convenient.
The use of eq.(\ref{eq:34-Lcoll-alt}) avoids the singularity given
by the form of eq.(\ref{eq:23-Lgcoll-tanaka}). Both parametrization
gives also a good agreement with numerical simulations. The choice
of one or the other is matter of a careful analysis of the experimental
results.

\end{document}